\documentclass[aps,pra,twocolumn,superscriptaddress,showpacs,showkeys,amsmath,amssymb]{revtex4}
\usepackage{amsfonts}
\usepackage{amssymb,amsmath}
\usepackage{mathrsfs}
\usepackage{latexsym}
\usepackage{amsmath}
\usepackage[cp1251]{inputenc}
\usepackage{graphicx}
\usepackage{dcolumn}
\usepackage{bm}
\usepackage{color}

\RequirePackage{ifthen}
\RequirePackage[pdfstartview=FitH]{hyperref}
\begin{document}
	
\title{Finite-temperature properties of low-dimensional bosons with three-body interaction}
\author{V.~Polkanov\footnote{e-mail: cogersum92@gmail.com}}	
\author{V.~Pastukhov\footnote{e-mail: volodyapastukhov@gmail.com}}
\affiliation{Professor Ivan Vakarchuk Department for Theoretical Physics, Ivan Franko National University of Lviv, 12 Drahomanov Street, Lviv, Ukraine}

	\date{\today}

	
	
\begin{abstract}
We discuss the finite-temperature properties of low-dimensional bosons with three-body interactions described by a Feshbach-resonance-like two-channel model. In particular, by using the approximate consideration that collects ring-like Feynman diagrams for the grand potential and resembles the three-body $t$-matrix approximation, we have computed the third virial coefficient, an equation of state, and the temperature depletion of the average number of closed-channel trimers. The calculated heat capacity demonstrates a non-monotonic temperature behavior, which is unusual for a low-dimensional Bose gas.
\end{abstract}
	
	\maketitle
\section{Introduction}
The ground-state behavior of low-dimensional bosons is known \cite{Pitaevskii_Stringari} to differ significantly from that at finite temperatures. The simplest illustrative example is an ideal Bose gas \cite{Ziff_77}, in which the Bose-Einstein condensate at absolute zero is completely destroyed due to Mermin-Wagner-Hohenberg theorem \cite{Mermin_Wagner,Hohenberg} by thermal fluctuations. A weak two-body repulsive interaction between Bose particles in lower dimensions typically ensures the superfluidity \cite{Popov} of the many-body system at absolute zero and provides additional stabilization at finite temperatures. The impact of three-body interactions on the properties of low-dimensional bosons at nonzero temperatures is less well studied. This problem is important for understanding few- and many-body physics in lower dimensions \cite{Mistakidis_et_al} because, in realistic setups, in addition to the two-body interaction, effective three-body (see \cite{Endov_25} for review) (and, in general, few-body) forces always arise \cite{Mazets_08} as a consequence of harmonic confinement in one or two spatial directions. These few-body effects are dominant \cite{Pricoupenko_Petrov_21} for two-body potentials with zero mean value. The residual three-body interaction emerges \cite{Valiente_19} between three bosons interacting through a pairwise potential with parameters fine-tuned to the two-body resonance. The enhancement of the three-body interaction originating from the atom-dimer scatterings can be realized \cite{Pricoupenko_Petrov_19} in a model of the narrow Feshbach resonance at small inter-channel couplings. The three-body interaction occurs \cite{Hammond_22} in Rabi-coupled spinor Bose condensates even at the mean-field level. Its emergence is related \cite{Tiengo_25} to the effective reduction of the degrees of freedom, but in the spin sector. This idea provides for the creation and control of few-body effective potentials in free space \cite{Petrov_14}, and on-site interactions in optical lattices \cite{Petrov_14_latt}.

In the last decade, low-dimensional systems with $\delta$-like three-body potential have attracted much attention in the literature. In general, the presence of this potential breaks the integrability \cite{McGuire_64} of the Lieb-Liniger model; however, an explicit analytic expression for the bound states of three bosons with contact two-body and three-body interactions is found \cite{Guijarro_et_al}. The system possesses two bound states, and the problem can be extended \cite{Nishida_18} to an arbitrary $N$-body sector, with a solution that requires extensive numerics. Neglecting the two-body attraction between bosons, Sekino and Nishida \cite{Sekino_18} solved the four-body problem and analyzed in detail the large-$N$ limit characterized by droplet-state formation. Exactly in two dimensions, the system of four bosons with a three-body interaction supports \cite{Nishida_17} the semisuper Efimov effect that switches to \cite{Hryhorchak_22,Polkanov_25} standard Efimov physics in fractional dimensions $1<d<2$. 

At least three fermionic species must be involved in a contact three-body interaction. The detailed study of the few-body states up to six particles in the system of $SU(3)$ fermions interacting through $\delta$-like three-body potential conducted by McKenney and Drut \cite{McKenney_19} revealed trimer formation in the system. If, however, one fermionic component is macroscopically occupied, the formation of a bipolaron \cite{Hryhorchak_26} is energetically preferred. Further generalization of the model in the unequal-mass limit and fractional dimensions leaves room \cite{Polkanov_24} for the $p$-wave Efimov effect in the four-body sector. From the effective field theory perspective, the three-body contact interaction is the most relevant perturbation in $d<2$, providing the quantum scale anomaly \cite{Drut_18} exactly in $d=1$. The collective behavior of one-dimensional $SU(3)$ fermions with three-body interaction was analyzed in the context of high-temperature virial expansion in \cite{Maki_19,Czejdo_20}, by means of Monte Carlo methods \cite{McKenney_20} and the three-body $t$-matrix approximation \cite{Tajima_24}. Except for vacuum-like trimers, the phase diagram of the macroscopic system includes an exotic Cooper-triple phase \cite{Akagami_21}, which remains stable \cite{Tajima_22} even at finite temperatures. In contrast, only the ground state of a many-boson system with a three-body $\delta$-potential is discussed in the literature. The universal relations and zero-temperature thermodynamics of a dilute homogeneous system were obtained in \cite{Pastukhov_19}. An effect of quantum anomaly in shifting frequencies of collective modes for a system loaded in a harmonic trap was elucidated in \cite{Valiente_Pastukhov}. Properties of the strongly-interacting bosons in quantum liquid and droplet states at absolute zero are studied in detail in Ref.~\cite{Morera_22}. The objective of the present article is to investigate the finite-temperature thermodynamics of a $d$-dimensional Bose gas with three-body interactions.

\section{Formulation}
\subsection{Two-channel model}
Consider a macroscopic number $N$ of bosonic atoms (of mass $m$) confined in a large volume $V=L^d$ with periodic boundary conditions applied. In the low-dimensional geometries ($d<2$), assuming that the two-body contact potential is fine-tuned to the unitary limit, the next relevant interaction is the three-body one. Above $d=1$, however, this interaction is not renormalizable. This particularly means that one can renormalize the three-body problem in higher dimensions by introducing the ultraviolet (UV) cutoff $\Lambda$ and simultaneously fine-tuning the three-body coupling; however, to make the four-, five-, and, in general, few-body sectors well-behaved, one necessarily needs to introduce higher-order interaction terms. An alternative approach to addressing this issue is to work from the outset with the two-channel model (recently introduced for fermions \cite{Polkanov_24} and bosons \cite{Polkanov_25}), which captures finite-range effects. The appropriate second-quantized Hamiltonian reads
\begin{eqnarray}\label{H}
&&H=\sum_{\bf p}\varepsilon_{\bf p}b^{\dagger}_{\bf p}b_{\bf p}+\sum_{\bf p}\left(\frac{\varepsilon_{\bf p}}{3}+\delta\omega_{\Lambda}\right)c^{\dagger}_{\bf p}c_{\bf p}\nonumber\\
&&+\frac{g}{3!L^d} \sum_{{\bf p}_1,{\bf p}_2,{\bf p}_3}(c^{\dagger}_{{\bf p}_1+{\bf p}_2+{\bf p}_3}b_{{\bf p}_1}b_{{\bf p}_2}b_{{\bf p}_3}+\textrm{h.c.}),
\end{eqnarray}
where the creation (annihilation) operators $b^{\dagger}_{\bf p}$ ($b_{\bf p}$) and $c^{\dagger}_{\bf p}$ ($c_{\bf p}$) of bosonic atoms and the closed-channel three-body bound states (trimers) satisfy the canonical commutation relations $[b_{\bf p}, b^{\dagger}_{{\bf p}'}]=[c_{\bf p}, c^{\dagger}_{{\bf p}'}]=\delta_{{\bf p},{\bf p}'}$ (with all other pairs commuting). The summation is carried out over the $d$-dimensional wave-vectors with the absolute values restricted by $\Lambda$, and $\varepsilon_{\bf p}=\frac{{\bf p}^2}{2m}$ is the single-particle dispersion relation. Being cutoff dependent, the detuning $\delta\omega_{\Lambda}=-g^2/(3!g_{3,\Lambda})$ realizes the renormalization in all few-body sectors. Constant \cite{Hryhorchak_22} $
g^{-1}_{3,\Lambda}=-\frac{1}{L^{2d}}\sum_{{\bf p}, {\bf p}'}\frac{1}{\varepsilon_{{\bf p}}+\varepsilon_{{\bf p}'}+\varepsilon_{{\bf p}+{\bf p}'}+|\epsilon_{\infty}|}$ is the broad-resonance ($g\to \infty$) bare coupling that provides a contact three-body interaction when the effective range $r_0$ of the microscopic three-body potential vanishes. The latter parameter $r_0$ is explicitly related $g\propto 1/(mr^{2-d}_0)$ to the inter-channel coupling $g$. When $g=0$, the Hamiltonian conserves the number of atoms $\sum_{\bf p}b^{\dagger}_{\bf p}b_{\bf p}$ and trimers $\sum_{\bf p}c^{\dagger}_{\bf p}c_{\bf p}$ separately. Any finite inter-channel coupling ($g\neq 0$) breaks this initial $U(1)\times U(1)$ symmetry to a single $U(1)$ group, whose generator is 
\begin{eqnarray}\label{N}
N=\sum_{\bf p}b^{\dagger}_{\bf p}b_{\bf p}+3\sum_{\bf p}c^{\dagger}_{\bf p}c_{\bf p}.
\end{eqnarray}
The Hamiltonian (\ref{H}) always possesses \cite{Polkanov_25} the three-body bound state with energy $\epsilon_{g}$ determined by the equation
\begin{eqnarray}\label{e_g}
\epsilon_{g}+\frac{g^2}{3!g_3}\left[1-\left(\frac{\epsilon_g}{\epsilon_{\infty}}\right)^{d-1}\right]=0,
\end{eqnarray}
in arbitrary $d$. Here, we introduced the `observable' coupling $g^{-1}_3=-\frac{\Gamma(1-d)}{(2\sqrt{3}\pi)^d}m^d|\epsilon_{\infty}|^{d-1}$, and the broad-resonance magnitude $\epsilon_{g\to \infty}\to\epsilon_{\infty}$ of the three-body binding energy. Although the observable coupling $g_3$ vanishes in the one-dimensional case, the three-body bound-state energy is finite $\epsilon_{g}=\frac{g^2m}{12\sqrt{3}\pi}\ln\left(\frac{\epsilon_g}{\epsilon_{\infty}}\right)$. The logarithmic behavior is intrinsic to 1D systems, with the contact three-body interaction manifesting the scale anomaly \cite{Drut_18}. Its origin is similar to that of two-dimensional systems with pairwise $\delta$-like interaction, because kinematically, these are the same problems. Nonzero effective range (finite $r_0$) presented in our model (\ref{H}) demolishes the anomalous contributions to the observables.

\subsection{Finite-temperature consideration}
It is well-known that bosons condense at absolute zero in any dimension $d>1$. Even in the one-dimensional case, bosons with weak repulsion should be superfluid, with a mechanism similar to the Berezinskii-Kosterlitz-Thouless transition. Thus, the ground state of our model is expected to be a complicated coexistence of two Bose condensates in higher dimensions, and the two-component superfluid exactly in $d=1$. This is true only for thermodynamic metastable states of the many-body system, which also possesses \cite{Nishida_18,Sekino_18} the non-thermodynamic (collapsed) $N$-body bound state that minimizes the total energy. Thermal fluctuations at any non-zero temperatures, however, break the superfluidity of the metastable ground states. From a practical point of view, it is therefore important to consider the effects of finite temperatures on the properties of bosons described by Hamiltonian (\ref{H}).

\begin{figure}[h!]
	\includegraphics[width=0.3\textwidth]{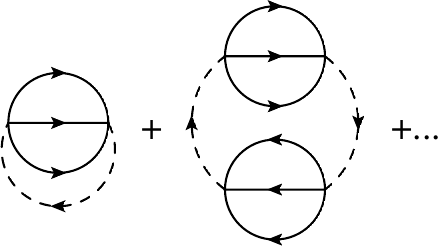}
	\caption{Infinite series of ring-like diagrams contributing to the thermodynamic potential $\Omega$. Solid and dashed lines represent bare propagators ($g=0$) of atoms and trimers, respectively.}
	\label{fig:rings}
\end{figure}

The grand thermodynamic potential $\Omega$ of the system in an ensemble with a fixed temperature $T$ (notation $\beta=1/T$ will also be used), volume $L^d$, and chemical potential $\mu$ can be formally calculated perturbatively in $g$. At zero order, the system is described by two ideal Bose gases with dispersions $\varepsilon_{\bf p}$ and $\frac{\varepsilon_{\bf p}}{3}$, respectively. To reproduce the simplest three-atom scatterings correctly, one needs to collect the series of ring-like diagrams presented in Fig.~\ref{fig:rings}.
In fact, every diagram of the series separately diverges at very low temperatures, independently of the inter-channel coupling $g$ magnitude. The sum of this series, however, is finite and together with the zero-order contribution results in
\begin{eqnarray}\label{Omega}
&&\Omega=\frac{1}{\beta}\sum_{\omega_n,{\bf p}}\ln\left[\xi_{\bf p}-i\omega_n\right]\nonumber\\
&&+\frac{1}{\beta}\sum_{\omega_n,{\bf p}}\ln\left[\xi_{c,{\bf p}}+\Sigma_{c,{\bf p}}(\omega_n)-i\omega_n\right],
\end{eqnarray}
(here $\xi_{\bf p}=\varepsilon_{\bf p}-\mu$, $\xi_{c,{\bf p}}=\frac{\varepsilon_{\bf p}}{3}-3\mu$, $\omega_n=2\pi n/\beta$ stands for the bosonic Matsubara frequency with $n$ being integer and factor $e^{i\omega_n0_+}$ is understood for regularization of sums) where the simplest self-energy insertion in the trimer propagator 
\begin{eqnarray}\label{Sigma_c}
\Sigma_{c,{\bf p}}(\omega_n)=\delta\omega_{\Lambda}-\frac{g^2}{3!}\Pi_{{\bf p}}(\omega_n),
\end{eqnarray}
is given by the elementary three-particle bubble, which, after evaluation of the Matsubara frequency summations, reads
\begin{align}\label{Pi_3}
\Pi_{{\bf p}}(\omega_n)=\frac{1}{L^{2d}}\sum_{{\bf p}_1+{\bf p}_2+{\bf p}_3={\bf p}}\frac{\prod_{i}(1+n_{{\bf p}_i})-\prod_{i}n_{{\bf p}_i}}{\xi_{{\bf p}_1}+\xi_{{\bf p}_2}+\xi_{{\bf p}_3}-i\omega_n},
\end{align}
with $n_{{\bf p}}=1/(e^{\beta \xi_{{\bf p}}}-1)$ standing for the Bose distribution of atoms.

With the thermodynamic potential (\ref{Omega}), one can obtain the temperature dependence of any observable in our system. First, however, we must express the chemical potential $\mu$ in terms of the total number of bosons $N$ (and temperature $T$) using the standard thermodynamic relation $N=-\frac{\partial\Omega}{\partial \mu}$. Important information about the system's composition is contained in the equilibrium number of trimers $N_c$. The occupation of trimers' energy levels, which is not fixed and depends on temperature, can be computed using an explicit expression for the trimer propagator
\begin{eqnarray}\label{N_c}
N_c=\frac{1}{\beta}\sum_{\omega_n,{\bf p}}\frac{e^{i\omega_n0_+}}{\xi_{c,{\bf p}}+\Sigma_{c,{\bf p}}(\omega_n)-i\omega_n}.
\end{eqnarray}
Within the same accuracy, the thermodynamic potential (\ref{Omega}) was obtained; it suffices to approximate the trimer self-energy using Eq.~(\ref{Sigma_c}). Indeed, by adopting the thermodynamic analog of the Hellmann-Feynman theorem $\langle \frac{\partial}{\partial g_{3,\Lambda}} H\rangle=\frac{\partial}{\partial g_{3,\Lambda}}\Omega$ (with $\mu$ and $T$ kept fixed), one readily obtains the identity $\frac{\partial}{\partial g^{-1}_{3,\Lambda}}\Omega=-\frac{g^2}{3!}N_c$. Then, taking into account expression (\ref{N_c}) for $N_c$ [with the self-energy (\ref{Sigma_c})], the integration over $g^{-1}_{3,\Lambda}$ of both sides yields the second term in the thermodynamic potential (\ref{Omega}). The first term of $\Omega$, in turn, appears as the integration constant at zero three-body coupling.

\begin{figure}[h!]
	\includegraphics[width=0.20\textwidth]{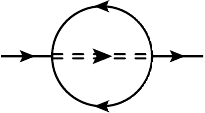}
	\caption{Simplest self-energy contribution to the propagator of bosons. The double-dashed line stands for the resummed trimer propagator [with the self-energy (\ref{Sigma_c}) inserted].}
	\label{fig:Sigma_b}
\end{figure}

An average number of composite bosons $N_c$ also determines Tan's contact parameter \cite{Tan,Braaten_2008}. For bosons with point-like three-body interaction, this parameter $\mathcal{C}_3$ was previously introduced in \cite{Pastukhov_19} for a one-dimensional system, and generalized to the arbitrary-$d$ case in Ref.~\cite{Hryhorchak_22}. Besides being a derivative of the thermodynamic potential with respect to observable three-body coupling $g_3$, the Tan contact determines the large-momentum tail of the particle distribution. Particularly, the summed over the Matsubara frequency bosonic correlator behaves like
\begin{eqnarray}\label{}
\frac{1}{\beta}\sum_{\omega_n}\frac{e^{i\omega_n0_+}}{\xi_{{\bf p}}+\Sigma_{b,{\bf p}}(\omega_n)-i\omega_n}\sim\mathcal{C}_3/|{\bf p}|^{4-d}.
\end{eqnarray}
at large $|{\bf p}|$s \cite{Hryhorchak_22}. Calculated in the adopted approximation, the self-energy $\Sigma_{b,{\bf p}}(\omega_n)$ of the atom's propagator is given by a single diagram presented in Fig.~\ref{fig:Sigma_b}.
Its consistency with the approximate thermodynamic potential (\ref{Omega}) can be proven by direct variational differentiation of $\Omega$ with respect to a bare atom's propagator. The leading-order asymptotics of the bosonic atoms' momentum distribution is easily obtained in the limit $|{\bf p}|\to \infty$, yielding a final result
\begin{eqnarray}\label{C_3}
\mathcal{C}_3=\mathcal{A}_d(mg)^2N_c/L^d,
\end{eqnarray}
(with dimensionless prefactor $\mathcal{A}_d=\frac{\Gamma(2-d/2)}{2(4\pi)^{d/2}(3/4)^{2-d/2}}$) for the three-body contact parameter. Despite its approximate derivation, formula (\ref{C_3}) is correct for any strength of the three-body coupling in the entire temperature region if one uses exact $N_c$. Note that the product $g^2N_c$ remains finite in the $g\to\infty$ limit, leaving Tan's contact (\ref{C_3}) well-defined even at broad resonance.

\section{Numerical results}
\begin{figure}[h!]
	\includegraphics[width=0.235\textwidth]{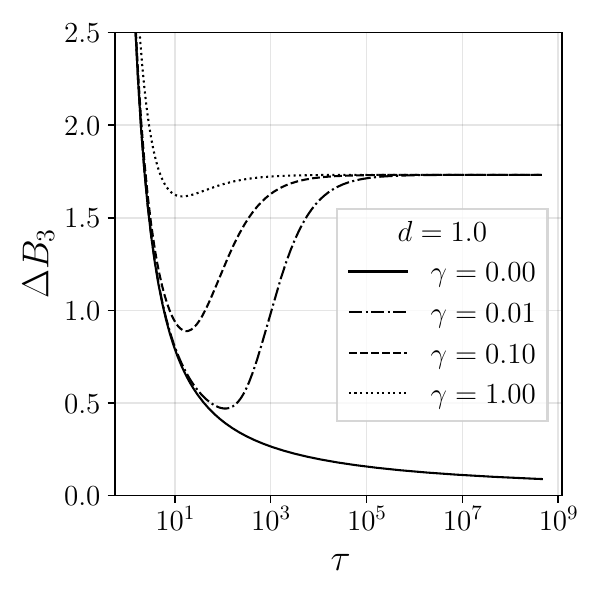}
    \includegraphics[width=0.235\textwidth]{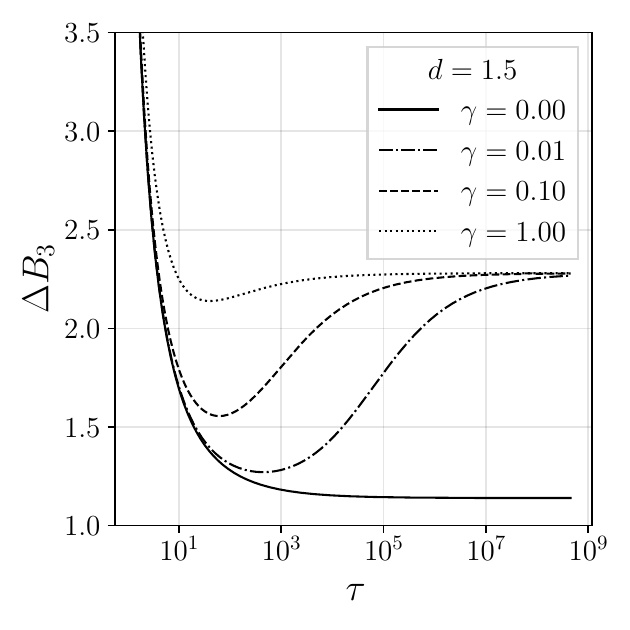}
	\caption{Temperature dependence ($\tau=T/|\epsilon_g|$) of the interaction-induced correction to the third virial coefficient in $d=1$ (left panel) and $d=1.5$ (right panel) at different inter-channel couplings $g$ parametrized by $\gamma=\ln(\epsilon_{\infty}/\epsilon_{g})$.}
	\label{fig:DeltaB_3}
\end{figure}

It is natural to start our discussion from the high-temperature region, where the system is almost classical with a large negative chemical potential. The latter fact allows the virial expansion in terms of the system's fugacity $e^{\beta\mu}$. The obtained thermodynamic potential (\ref{Omega}) guarantees the correct low-density series for pressure $p=-\Omega/L^d$
\begin{eqnarray}\label{p_virial}
p=\frac{T}{\lambda^d}\left[B_1e^{\beta\mu}+B_2e^{2\beta\mu}+B_3e^{3\beta\mu}+\dots\right],
\end{eqnarray}
(with $\lambda=\sqrt{2\pi/mT}$ being the thermal de Broglie wavelength) up to the third virial coefficient. Furthermore, in the high-temperature limit, we can simplify the trimer self-energy by neglecting Bose distributions in (\ref{Pi_3}). For the system with only three-body interaction, the first two virial coefficients are trivial, $B_1=1$, $B_2=1/2^{d/2+1}$, while the third one, $B_3=1/3^{d/2+1}+\Delta B_3$, is modified by interaction. We have plotted in Fig.~\ref{fig:DeltaB_3}
a typical behavior of the interaction-induced effect, $\Delta B_3$, on the third virial coefficient. The temperature is measured in units of modulus three-body binding energy $|\epsilon_{g}|$, and the distance to the broad-resonance zero-range (and appropriately $g^{-1}=0$) regime is parametrized by $\epsilon_{g}/\epsilon_{\infty}=e^{-\gamma}$. It is seen that the behavior of $\Delta B_3$ at any finite range ($\gamma\neq 0$) is completely different from the broad-resonance case \cite{Maki_19,Czejdo_20}. This can serve as a potential experimental test for uncovering the finite-range three-body interaction.

The first step in calculations at finite temperatures is to rewrite the chemical potential in terms of the system's average density $N/L^d$. To simplify numerics, we neglected all $n_{{\bf p}}$s in the three-particle bubble $\Pi_{{\bf p}}(\omega_n)$, determining the trimer self-energy (\ref{Sigma_c}). A similar approximation has recently been adopted \cite{Tajima_25} for finite-temperature $SU(3)$ fermions. This, however, has no effect on the final result in almost all temperature regions. The reason for this lies in the behavior of the chemical potential. At high temperatures, it suppresses the Bose distribution by a small factor $e^{\beta\mu}$. At the same time, in the low-temperature region, the thermally-excited bosons are almost absent in the system due to the energy gap $|\epsilon_g|/3$ in a single-particle spectrum provided by the three-body bound state. Within this approximation, the second term in (\ref{Omega}) reproduces the thermodynamic potential of an ideal Bose gas formed by vacuum-like trimers.

\begin{figure}[h!]
	\includegraphics[width=0.235\textwidth]{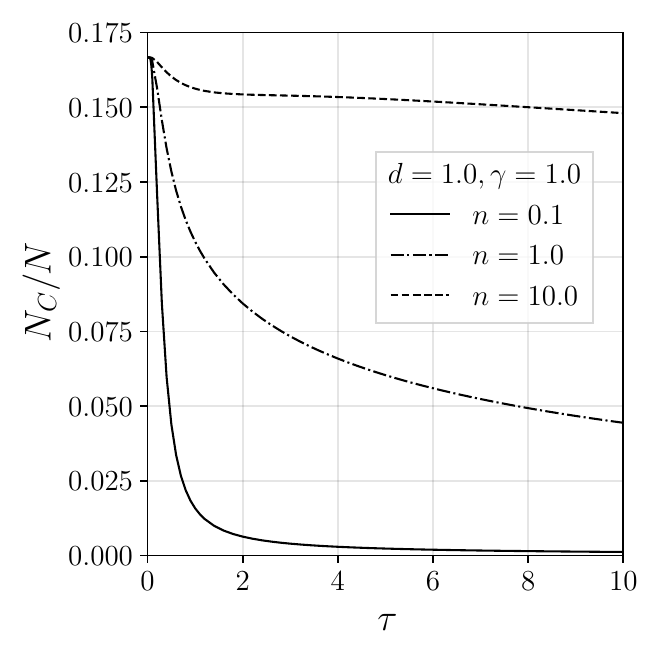}
    \includegraphics[width=0.235\textwidth]{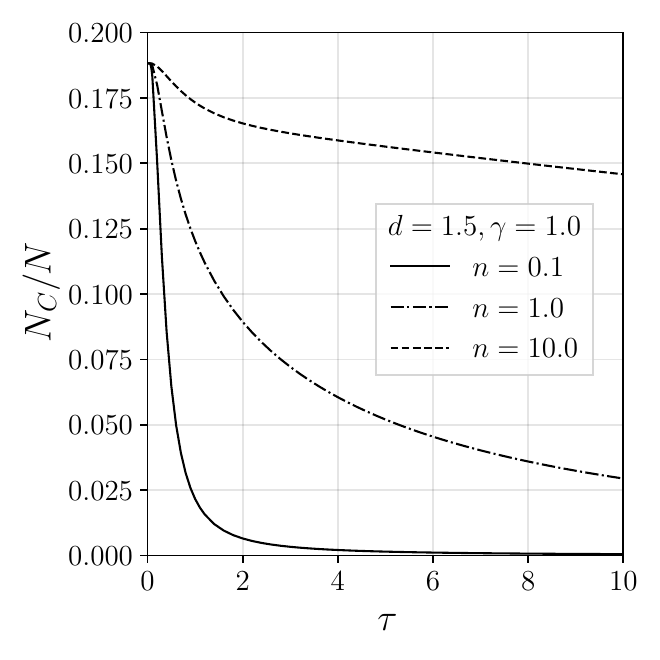}
	\caption{Temperature ($\tau=T/|\epsilon_g|$) depletion of the closed-channel trimers at three dimensionless densities ($n=(N/L^d)(2\pi/m|\epsilon_g|)^{d/2}$) and inter-channel coupling $\gamma=\ln(\epsilon_{\infty}/\epsilon_{g})=1$ in $d=1$ (left panel) and $d=1.5$ (right panel).}
	\label{fig:n_c}
\end{figure}

With the numerically computed temperature dependence of the chemical potential, one can calculate all observables. In the following, we continue to focus on two examples of spatial dimensionality, namely, $d=1$ and $d=1.5$. Other fractional dimensions demonstrate qualitatively similar behavior of thermodynamic functions to that of the $d=1.5$ case. At first, we calculated (see Fig.~\ref{fig:n_c})
the temperature depletion of the closed-channel trimer in the system. Their number is maximal in the ground state, independent of the density [measured in Fig.~\ref{fig:n_c} in dimensionless units $n=(N/L^d)(2\pi/m|\epsilon_g|)^{d/2}$], and is fully determined by the quasiparticle residue of composite particles. At higher temperatures, the number of closed-channel trimers decreases monotonically and asymptotes to zero as $T\to \infty$. A similar temperature behavior is expected for the three-body contact parameter (\ref{C_3}).

\begin{figure}[h!]
	\includegraphics[width=0.235\textwidth]{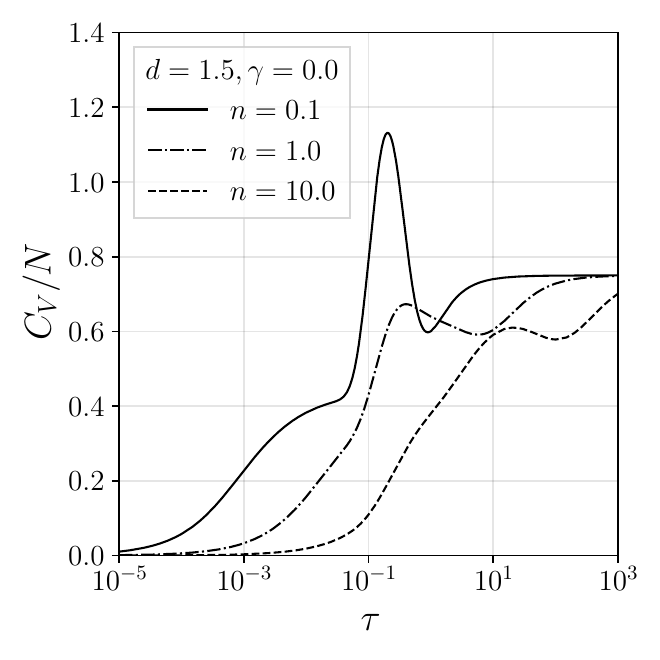}
    \includegraphics[width=0.235\textwidth]{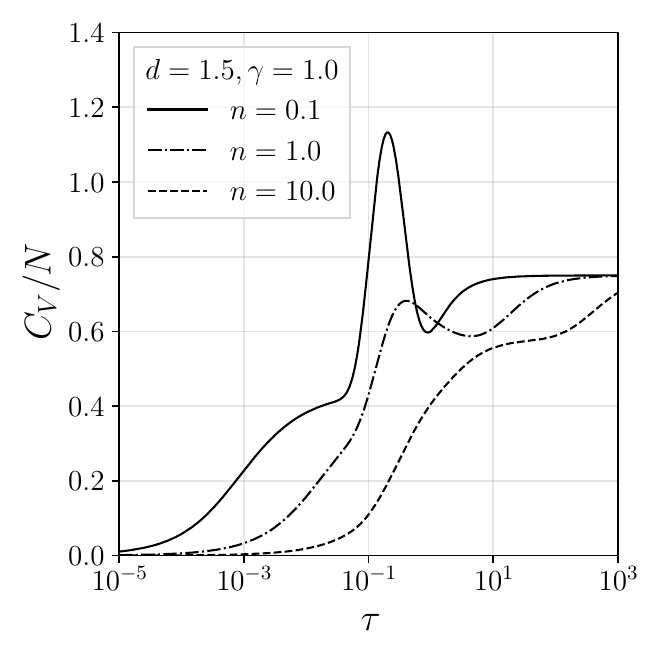}
	\caption{Isochoric heat capacity per particle as a function of temperature (in units of $|\epsilon_g|$) in $d=1.5$ at different densities (in dimensionless units) without (left panel) and with (right panel) inclusion of finite-range effects.}
	\label{fig:cap_15}
\end{figure}

\begin{figure}[h!]
	\includegraphics[width=0.235\textwidth]{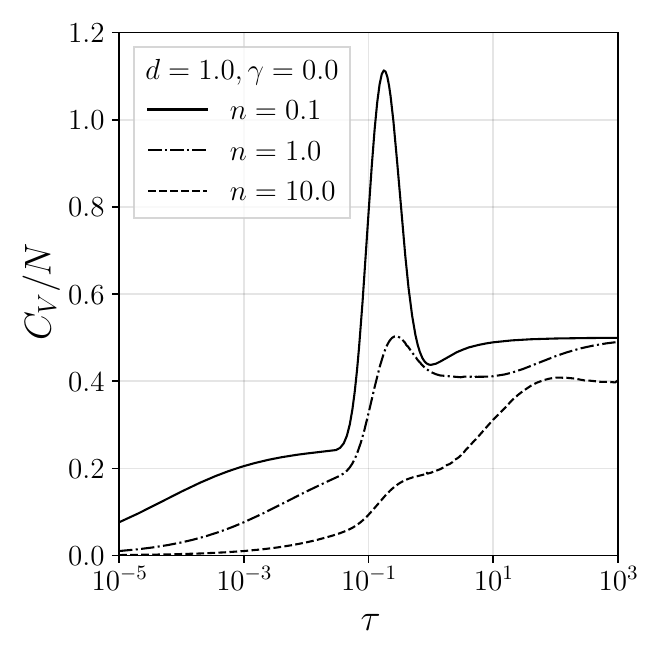}
    \includegraphics[width=0.235\textwidth]{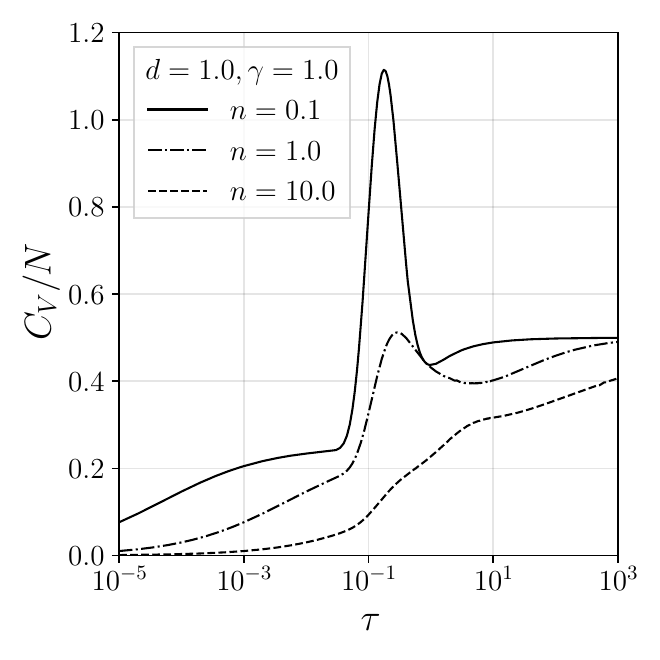}
	\caption{Temperature behavior of the specific heat in one dimension. Left (right) panel displays results at different densities without (with) inclusion of finite-range effects.}
	\label{fig:cap_10}
\end{figure}

Calculating the system's isochoric heat capacity is more tedious. In the grand-canonical ensemble, we first obtain the entropy of the system by direct differentiation $S=-\left(\frac{\partial\Omega}{\partial T}\right)_{V,\mu}$ (this can be done analytically within the adopted approximation), and then numerically calculate the derivative $C_V=T\left(\frac{\partial S}{\partial T}\right)_{V,N}$ by the finite-difference method. Technically, it requires more computational time because the equation determining the chemical potential must be solved twice at two nearby temperatures at each point. Results for $C_V$ per particle are presented for $d=1.5$ and $d=1$ in Fig.~\ref{fig:cap_15} and Fig.~\ref{fig:cap_10}, respectively.
The obtained curves of the specific heat clearly demonstrate a non-monotonic behavior with temperature, which is not intrinsic \cite{Ziff_77,Rovenchak_15,Li_15} for the low-dimensional ($d\le 2$) non-interacting bosons. Independent of spatial dimension, the effect is more pronounced at lower densities and closer to the broad resonance (smaller effective ranges). To understand non-monotonicity, one should keep track of the $N_c$ (see Fig.~\ref{fig:n_c}) behavior. Indeed, a peak of $C_V$ can be associated with the abrupt decrease in the number of composite bosons. The latter also applies to the open-channel trimers. This immediately increases the number of degrees of freedom (by two for each process of trimer dissociation), thereby maximizing the heat capacity of the system. In general, we can conclude that any substantial changes in the behavior of the derivative $\left(\frac{\partial N_c}{\partial T}\right)_{V,N}$ (obviously, the same applies to the number of open-channel three-body bound states) are reflected in the $C_V$ temperature dependence.

\begin{figure}[h!]
	\includegraphics[width=0.235\textwidth]{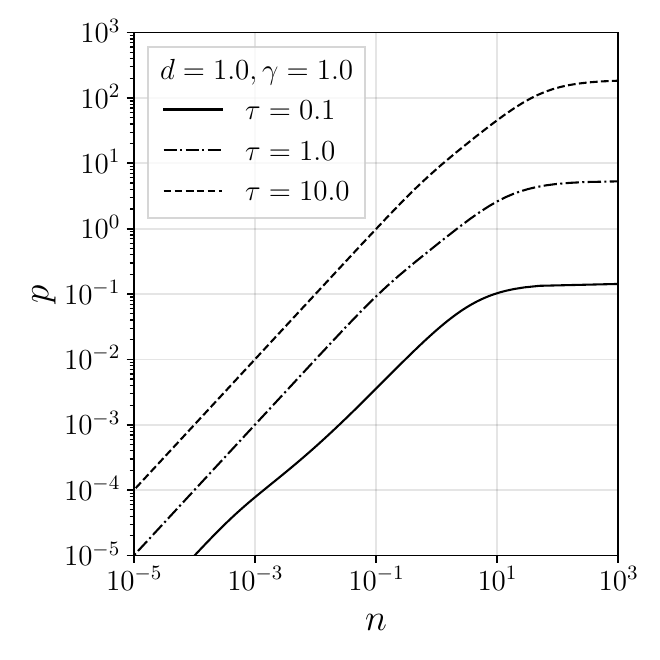}
    \includegraphics[width=0.235\textwidth]{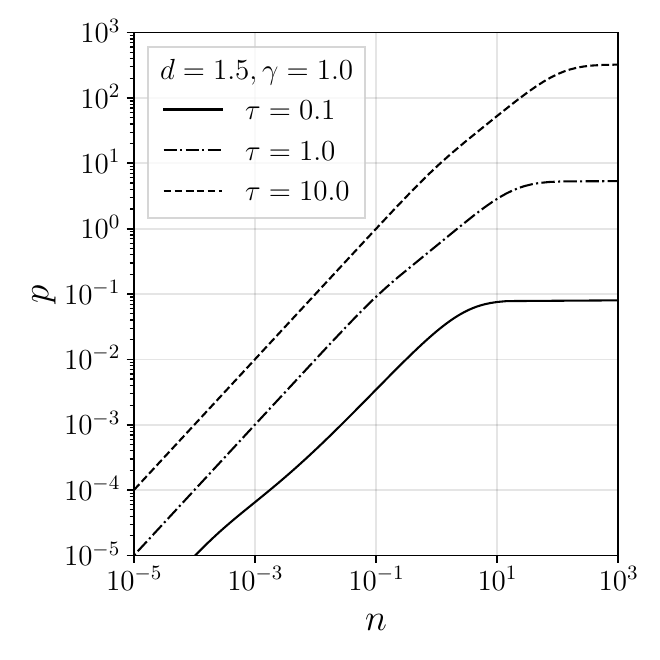}
	\caption{Examples of isotherms in dimensionless ($p,n$) variables for several temperatures (in units of $|\epsilon_g|$) and inter-channel coupling $\gamma=\ln(\epsilon_{\infty}/\epsilon_{g})=1$ in $d=1$ (left panel) and $d=1.5$ (right panel). The curves are qualitatively the same for all other $\gamma$s, including $\gamma=0$ case.}
	\label{fig:p_n}
\end{figure}

Further information about the system's phase diagram can be obtained from its equation of state. In this case, we solved the equation for the chemical potential at a few fixed temperatures by varying the system's total density. In Fig.~\ref{fig:p_n},
we plotted isotherms of bosons with a three-body interaction as functions of the dimensionless density in $d=1.0$ (left panel) and $d=1.5$ (right panel). Numerical computations revealed a weak dependence of the isotherm form on the effective range of the potential. Some quantitative differences were found only at high temperatures. A low-density regime clearly demonstrates classical ideal-gas behavior $p\propto n\tau$ that is altered by the classical-quantum crossover region ($n\sim 10^{-1}$ at $\tau=0.1$) and ultimately transitions to fully quantum behavior in the high-density limit. Notably, the inequality $\left(\frac{\partial p}{\partial n}\right)_{T}>0$ is always satisfied, signaling the thermodynamic stability of bosons with three-body interaction. At high densities (high degeneracy temperatures), this derivative tends to zero, which is typical behavior for the compressibility of an ideal Bose gas at low temperatures.

\section{Summary}
In conclusion, we have studied the finite-temperature thermodynamics of a low-dimensional Bose gas with three-body interactions. Adopting the two-channel model introduced earlier \cite{Polkanov_25}, which includes the effects of the finite ranges of the realistic inter-boson potential, we calculated the grand potential of the system, accounting for an infinite series of ring-like Feynman diagrams. At first glance, the obtained thermodynamic potential in this approximation can be thought of as the one describing two ideal Bose gases: the gas of bosonic atoms and the gas of trimers, respectively. This suggests somewhat trivial thermodynamics. However, the calculated properties of the system are very different from those of low-dimensional non-interacting bosons. In particular, we found a non-monotonic dependence of the system's heat capacity on temperature, which is intrinsic to the Bose gas (both ideal and interacting) in higher dimensions. A key reason for the non-monotonicity lies in the intense thermal dissociation of the three-body bound states of bosonic atoms, both in the open and closed channels. The temperature dependence of the number of closed-channel trimers, which can be explicitly calculated, just illuminates this decay process. Furthermore, in the limit of broad resonance, where there are no composite particles (but three-boson bound states still exist), the heat-capacity non-monotonicity is more pronounced. 

An important outcome of the equation-of-state calculations is the thermodynamic stability of bosonic atoms with three-body interactions, at least at finite temperatures. This can be explicitly shown by analyzing the system's isothermal compressibility. Finally, we have identified the impact of finite ranges of inter-particle potential on various observables. The most drastic difference between the zero-range and finite-range regimes is seen in the temperature dependence of the third virial coefficient.

\begin{center}
	{\bf Acknowledgements}
\end{center}
We thank Prof. A.~Rovenchak and Dr. O.~Hryhorchak for the discussion of results. The work of V. Polkanov was supported by Project FF-29Nf (No.~0126U001581) from the Ministry of Education and Science of Ukraine.

\end{document}